\pdfoutput=1
\documentclass{agujournal}

\usepackage{natbib,color,graphicx}
\usepackage{textcomp}
\usepackage{float}
\usepackage{soul}
\usepackage{sidecap}
\usepackage{subfig}
\usepackage{enumerate}
\usepackage{amsmath}

\usepackage{textpos}

\journalname{JGR-Space Physics}

\begin{document}

%
%

\title{Simulation of Focusing Effect of Traveling Ionospheric Disturbances on Meter-Decameter Solar Dynamic Spectra}

\begin{textblock*}{100mm}(.19\textwidth,-2.6cm)
\textsc{Accepted by JGR-Space Physics, October 2018}
\end{textblock*}

%
%

\authors{Artem Koval\affil{1}, Yao Chen\affil{1}, Aleksander Stanislavsky\affil{3}, Anton Kashcheyev\affil{2,3},  Qing-He Zhang\affil{1}}

\affiliation{1}{Shandong Provincial Key Laboratory of Optical
Astronomy and Solar-Terrestrial Environment, \\ and Institute of
Space Sciences, Shandong University, Weihai, China.}
\affiliation{2}{Abdus Salam International Centre for Theoretical Physics, Trieste, Italy.}
\affiliation{3}{Institute of Radio Astronomy of National Academy
of Sciences of Ukraine, Kharkiv, Ukraine.}


\correspondingauthor{Atem Koval}{art$\makebox[0.1cm]{\hrulefill}$koval@yahoo.com}


\begin{keypoints}
\item Simulation of the focusing effect of TIDs on solar radiation showing up in dynamic spectra, in particular, the Spectral Caustics (SCs)
\item Morphological analysis of the spectral perturbations obtained by the simulation
\item Interpretation of the observed seasonal dependency in the SC occurrence rate
\end{keypoints}

\begin{abstract}
For the first time we present simulation results of the focusing effect of the ionospheric plasma density irregularities, namely, Medium Scale Traveling Ionospheric Disturbances (MSTIDs), on solar radio emission by applying a ray-tracing method to the Earth's ionosphere with MSTIDs. With this technique we investigate the focusing effect which manifests itself in the form of peculiar spectral disturbances in intensity with specific morphology, so-called Spectral Caustics (SCs), occasionally appearing in dynamic spectra of different solar radio instruments operating in the meter-decameter wavelength range. We show that the simulated spectral shapes of SCs are in good agreement with the ones detected in real solar radio spectrograms. In particular, SCs that are classified as inverted V-like, V-like, X-like, and fiber-like types have been reproduced. It is also found that the seasonal dependence in the occurrence of SCs, which has been discovered recently, can be understood through a strong relation between the focusing frequency - the most important characteristic point in most SCs patterns - and the elevation angle of the Sun. We find that under typical parameters of MSTIDs with spatial and temporal periods set to be 300 km and 40 min, respectively, the focusing frequency decreases with the growth of the elevation angle. Physical interpretations of the results and implications on the analysis of solar radio data with SCs are discussed.
\end{abstract}

\section{Introduction}
\label{Sect:Introduction}

Traveling Ionospheric Disturbances (TIDs) represent one particular type of the Earth's ionosphere irregularities. They are the ionospheric response to the passage of Acoustic Gravity Waves (AGWs) and represent wave-like electron density structures propagating in the ionosphere \citep{Hines1960}. It is believed that AGWs could be generated by various natural or anthropogenic processes (earthquakes, hurricanes, moving solar terminator, auroral activity, powerful explosions, missile launches, ionosphere modification tests, etc.) \citep{Galushko1998, Fedorenko2013}. TIDs are classified according to their spatial and temporal scales (wavelengths and periods) and horizontal velocities. Thus, TIDs with spatial (>1000 km) and temporal ($\sim$0.5-3 h) periods belong to the large-scale class (LSTIDs), while those with spatial and temporal scales of 100-600 km and 0.25-1 h, respectively, are classified as the medium scale class (MSTIDs) \citep{Hunsucker1982}. It is believed that TIDs can propagate in free wave-mode, being reflected by the Earth's surface, as well as in guided wave-mode \citep{Fedorenko2013}. In the latter case, AGWs/TIDs move parallel to the ground surface at heights 250-300 km in a slab. This directed motion modulates the electron density distribution in space. It leads to a modification of plasma parameters, namely the refractive index, and affects the propagation of radio waves through the slab. The presence of TIDs may cause a considerable inaccuracy in satellite navigation positioning using electromagnetic waves at GHz range (see, e.g., \citet{Hernandez2006} and references therein). In addition, the variations of plasma parameters also strongly affect the lower-frequency electromagnetic waves that can result in focusing or amplification of the incident radiation in particular cases.

The effect of TIDs on solar radio emission recorded on the ground was first noticed during observations with the Nan\c{c}ay Radio Heliograph (NRH) at meter wavelengths \citep{Bougeret1973, Bougeret1981}. The quasi-periodic variations of apparent positions of solar radio sources were interpreted as pure refraction induced by MSTIDs. These works stimulated further efforts to establish the basic concepts of the phenomena by using a model to simulate the disturbing effects of TIDs on solar radio emission \citep{Meyer-Vernet1980, Meyer-Vernet1981}. The theoretical treatment is given in terms of focusing and diffraction of an incident radio wave by ionospheric irregularities that are represented as a one-dimensional sinusoidal phase screen. It was shown that the phase changes induced by TIDs can give rise to specific intensity perturbations to the solar dynamic spectra observed in meter-decameter wavelengths. These spectral perturbations are known as Spectral Caustics (SCs) due to their morphology as well as their formation mechanism.

Following studies on the issue by \citet{Genova1983, Mercier1986a, Mercier1986b, Mercier1989} used spectral data obtained by the Nan\c{c}ay Decameter Array (NDA) or/and imaging data with the NRH \citep{Lecacheux2000}. The NDA works within 10-80 MHz range. Despite an initial rapid progress in the study of the phenomena, follow-up investigations have been suspended until recently. Lately, \citet{Koval2017} have revived the subject by carrying out a long-term statistical analysis of SC occurrence using the daily NDA dynamic spectra for 17-year period (1999-2015 years). The SCs were detected confidently in 129 observational sessions. For the first time the authors have introduced a classification of the SCs and divided them into several particular types, namely inverted V-like, V-like, X-like, fiber-like, and fringe-like, according to their spectral morphology (for details see \citep{Koval2017}). Most SCs can be distinguished from well-known solar radio bursts due to their sharp edges (i.e., rapid enhancement in intensity) in the dynamic spectra. Besides, it has been revealed that the SCs have strong seasonal and solar-cycle dependencies. In particular, the occurrence rate of SCs is higher in autumn-winter period ($\approx95\%$ of days with the SCs) than in spring-summer period ($\approx5\%$ of days with the SCs). This was attributed to the TID occurrence rate that shows similar seasonal variations as demonstrated in many papers \citep{Hernandez2006, Otsuka2013}.

The presence of SCs in solar dynamic spectra can be explained in terms of deviations of the refraction angle which are proportional to the electron density gradient along the wave propagation direction. For electron density irregularities of parabolic or close-to-parabolic spatial shapes, the refraction angle changes linearly with the distance away from the center of the irregularity. This may converge the rays to a point at a certain frequency, namely the focusing frequency. The distance from this focusing point to the density irregularity can be interpreted as the focal length and the irregularity becomes a plasma lens. At frequencies lower than the focusing frequency, one expects a kind of caustic surfaces in three-dimensional space that are produced by rays diffracted by the plasma lens.

With the above statement, it is easy to understand why TIDs can have a focusing effect on solar radiation. In addition, the wave-like TID structure represents an alternating sequence of cells with rising and falling magnitudes of electron density. The refractive index for an electromagnetic wave takes a form $n = \sqrt{1 - 80.6\cdot10^{6}\cdot N_{e}/f^{2}}$, where $N_{e}$ and $f$ are electron density (in $\textrm{cm}^{-3}$) and radiation frequency (in Hz), respectively. It means that the deviations would be directed towards the center of that TID's cell which has low magnitude of electron density (high value of the refractive index) and is sandwiched between two cells with high magnitudes of electron density (low values of the refractive index). Therefore, TIDs are able to act on solar radiation as periodic convergent plasma lenses. Thus, the Spectral Caustics, being captured in a time-frequency plane (i.e., a dynamic spectrum), reflect the manifestations of the spatial-temporal caustic surfaces. Note, a SC consisting of two (front and back) envelopes crossing at the focusing frequency is rather an ideal case. Because of peculiar conditions in the solar radiation as well as dynamics and structures of the ionosphere, the focusing effect of TIDs on solar radio emission results in a wide variety of different spectral shapes of SCs, which have been categorized into several kinds, as mentioned.

Up to the recent time, there has been only one proposed model of SCs \citep{Meyer-Vernet1980, Meyer-Vernet1981}. It is based on the application of the theory of phase screen \citep{Rino1979}. The SCs are calculated as a result of the Fresnel diffraction of incident radiation on this screen. In this paper, we approach the issue of SCs modeling by using geometrical optics. Thus, we make a novel effort to apply the ray tracing method to calculate the trajectories of solar radio rays through the disturbed terrestrial ionosphere, including MSTIDs, described by appropriate electron concentration model. In this way we examine the emergences of SCs in solar dynamic spectra. We also analyze the reason for the pronounced seasonal dependence in the SC occurrence discovered in solar radio observations at middle latitudes in Europe. This work offers a novel and more realistic approach to the problem of the SC generation during solar radio observations.

The present paper is organized in the following way. The daytime MSTID model and the ray-tracing algorithm will be described in Sections~\ref{Sect:Day-time MSTID model} and \ref{Sect:Ray tracing method}, respectively. The simulation results will be presented in Section~\ref{Sect:Simulation Results}, and further discussed in Section~\ref{Sect:Discussions}. The main conclusions will be given in Section~\ref{Sect:Conclusions}.

\section{Daytime MSTID model}
\label{Sect:Day-time MSTID model}

To date, several TID models have been established and presented in the literature. In particular, \citet{Leitinger2005} developed a model for simulation of large-scale TIDs with several electron density background models. The authors demonstrated the applications of different background models like the IRI, the NeQuick, and others in the modelling of TIDs. Recently \citet{Fedorenko2013} have designed a multiscale semi-empirical model of TIDs using their long-term measurements with trans-ionospheric sounding. Besides, the model was validated by reproducing/simulating available data on plasma density perturbations generated by different sources (e.g., nuclear and volcano explosions, earthquake, energetic proton precipitation events). In our study, to simulate the daytime MSTIDs caused by a gravity wave, we refer to the model given by \citet{Hooke1968} that has been used by many ionosphere physicists for years \citep{Fukao1993,Shiokawa2003,Otsuka2013}.

It is well known that an AGW represents an oscillating parcel of neutral air that propagates in space. Due to collisions between neutral air particles and ions, the ions are set in motion along the geomagnetic field. According to \citet{Hooke1968}, this leads to the following plasma density perturbation
\begin{equation}
N_{e}^{\prime} = i\omega^{-1}u_{||}\left[\frac{\partial N_{{e}_{0}}}{\partial z}\textrm{sin}I-ik_{||}N_{{e}_{0}}\right],
\label{Eq1}
\end{equation}
where $i$ is imaginary unit, $\omega$ is angular frequency of a gravity wave, $u_{||}$ and $k_{||}$ represent the neutral wind velocity and the wave vector of a gravity wave in the direction of the geomagnetic field, respectively, $I$ is the dip angle (magnetic inclination) of the geomagnetic field, and $N_{{e}_{0}}$ is the background electron density only depending on the altitude in the $z$-direction. For the simulations, the background electron density profile is taken to be the Chapman function \citep{Francis1974}:
\begin{equation}
N_{{e}_{0}} = N_{\textrm{m}}\,\textrm{exp}\frac{1}{2}\left[1 - \frac{z-z_{\textrm{m}}}{H}-\textrm{exp}\left(-\frac{z-z_{\textrm{m}}}{H}\right)\right],
\label{Eq2}
\end{equation}
where $N_{\textrm{m}}$ is peak electron density, $z_{\textrm{m}}$ is height of peak electron density, and $H$ is scale height.

The horizontal component of the wave number $k_{x}$ is obtained through the specified horizontal wavelength $\lambda_{x}$. The linear dispersion relation of a gravity wave relates $k_{x}$ with the vertical component of the wave number $k_{z}$ in the following form \citep{Hines1960,Shiokawa2003}:
\begin{equation}
k_{z}^{2} = \frac{\omega_{b}^{2}}{(U-c)^2}-k_{x}^{2}-\frac{1}{4H^2},
\label{Eq3}
\end{equation}
where $\omega_{b}$, $U$, $c$, and $H$ are Brunt-V$\"{a}$is$\"{a}$l$\"{a}$ or buoyancy angular frequency, background wind velocity in the direction of wave propagation, wave phase velocity, and scale height, respectively. Hence the wave vector along the direction of the geomagnetic field for Equation~(\ref{Eq1}) can be calculated as
\begin{equation}
k_{||} = \sqrt{k_{x}^{2} + k_{z}^{2}}\cos\left(\arctan\left(\frac{k_{z}}{k_{x}}\right)-I\right)\cos([180-A]\pm\delta),
\label{Eq4}
\end{equation}
where $A$ is azimuth of a gravity wave, $\delta$ is magnetic declination. The argument of the first cosine represents a difference between the slope angle of a gravity wave in the vertical plane and magnetic inclination, while the argument of the second cosine signifies the difference between azimuth angle and magnetic declination in the horizontal plane.

Consequently, the ionospheric electron density that includes fluctuations induced by AGW can be approximated as
\begin{equation}
N_{e} = N_{{e}_{0}} + \gamma |N_{e}^{\prime}|\cos(\omega t - k_{x}x -k_{z}z),
\label{Eq5}
\end{equation}
where the coefficient $\gamma$ can be used to adjust the amplitude of the electron density perturbation relative to the ambient electron density.

The numerical values of the parameters in the above expressions and details of the MSTID model will be given in Section~\ref{Sect:Simulation Results}.

\section{The ray-tracing method of radio wave propagation}
\label{Sect:Ray tracing method}

To simulate a propagation of radio rays in the modelled ionosphere, as described above, we have to trace a beam trajectory in a nonuniform medium. In solar radio astronomy, various quite sophisticated approaches have been used for this purpose \citep{Thejappa2007,Benkevitch2010}. However, there exists a simple algorithm based on the piecewise linear approximation of the smooth trajectory of a beam, in which the solar atmosphere is divided into layers \citep{Newkirk1961,Stanislavsky2013,Stanislavsky2016}, and the direction of the refracted beam is found with the Snell's law. We adopt this simple algorithm for our study, whose details will be introduced below with respect to the analysis of the radio rays (or beams) tracing through the Earth's ionosphere. Note that we will consider two-dimensional problem.

Following the basic principles of geometric optics \citep{Born1965}, the equation of a beam propagating in a medium with changing refractive index can be represented by:
\begin{equation}
\vec{S}_{2} = \frac{n_1}{n_2}\vec{S}_{1}-k\vec{Q},
\label{Eq6}
\end{equation}
where $\vec{S}_{1}$ and $\vec{S}_{2}$ are the unit vectors of incident and refracted beams, respectively, $\vec{Q}$ is the unit normal vector, $n_1$ and $n_2$ are the refractive indexes of media corresponding to the incident and refracted beams, respectively; the coefficient $k$ will be determined below. The direction of the straight line in a two-dimensional space can be specified by two vectors ($\vec{S}_{1}$ and $\vec{S}_{2}$) in the following form
\begin{equation*}
\vec{S}_{1} = a_{1}\vec{x}+b_{1}\vec{y},\qquad
\vec{S}_{2} = a_{2}\vec{x}+b_{2}\vec{y},
\end{equation*}
where $\vec{x}$, $\vec{y}$ are unit vectors along the axes of the Cartesian rectangular coordinates, and $a_{i}$, $b_{i}$ are the direction cosines. The unit normal vector $\vec{Q}$ can be expressed in the same way:
\begin{equation*}
\vec{Q} = a_{Q}\vec{x}+b_{Q}\vec{y},
\end{equation*}
which can be calculated by the change of the gradient of the electron density of the plasma. The changes of the electron concentration along $x$ and $y$ are described by
\begin{equation}
\Delta N_x = N_e(x+\delta x, y) - N_e(x-\delta x, y),\qquad
\Delta N_y = N_e(x, y+\delta y) - N_e(x, y-\delta y),
\label{Eq7}
\end{equation}
where the point ($x$,$y$) corresponds to the location where the refraction takes place. Then we get
\begin{equation*}
\begin{aligned}
(\Delta N)^2 = (\Delta N_x)^2 + (\Delta N_y)^2,\\
D = (\Delta N_x)a_1 + (\Delta N_y)b_1.
\end{aligned}
\end{equation*}
This allows us to determine the direction cosines
\begin{equation*}
a_{Q} = -\frac{\Delta N_x}{\Delta N}\frac{D}{|D|},\qquad
b_{Q} = -\frac{\Delta N_y}{\Delta N}\frac{D}{|D|}.
\end{equation*}
To find the coefficient $k$, Equation~(\ref{Eq6}) should be multiplied by $\vec{Q}$ to further calculate its scalar product using well-know formula for stratified medium $n_1/n_2 = \sin\theta_2/\sin\theta_1$, where $\theta_1$ and $\theta_2$ are angles of incidence and refraction, respectively. Consequently, an expression for $k$ is obtained,
\begin{equation}
k = \cos\theta_2 + \frac{n_1}{n_2}\cos\theta_1 = \frac{n_1}{n_2}(\vec{S}_{1}\cdot\vec{Q}) + \textrm{sgn}(n_2-n_1)\sqrt{1+\left(\frac{n_1}{n_2}\right)^2(\vec{S}_{1}\cdot\vec{Q})^2 - \left(\frac{n_1}{n_2}\right)^2},
\label{Eq8}
\end{equation}
where the signum function defines the sign of the difference of refractive indices. Equations (\ref{Eq6}) and (\ref{Eq8}) cannot be used in the turning point of a beam, because the reflection takes place there. At that point $\vec{S}_{1} = -\vec{S}_{2}$, $n_1\rightarrow n_2$, and so $k = 2(\vec{S}_{1}\cdot\vec{Q})$ which can be substituted into Equation~(\ref{Eq6}). The refractive indices $n_1$ and $n_2$ are found from expressions
\begin{equation}
n_1 = \sqrt{1-(f_p(x+\delta x,y+\delta y)/f)^2}, \qquad
n_2 = \sqrt{1-(f_p(x-\delta x,y-\delta y)/f)^2}.
\label{Eq9}
\end{equation}
By specifying the direction of $\vec{S}_1$ through its direction cosines $a_1, b_1$ and computing the direction cosines of the normal vector $a_Q, b_Q$ at the refraction point, from Equation~(\ref{Eq6}) we can determine the direction cosines of the refracted vector $\vec{S}_2$:
\begin{equation*}
a_{2} = \frac{n_1}{n_2}a_1-ka_{Q}, \qquad
b_{2} = \frac{n_1}{n_2}b_1-kb_{Q}.
\end{equation*}
Thus, in an iterative way a ray extends from its current position $(x_{i},y_{i})$ with direction cosines $a_1, b_1$ to a new one $(x_{i+1},y_{i+1})$:
\begin{equation}
x_{i+1} = x_{i}+\Delta s\,a_{2},\qquad
y_{i+1} = y_{i}+\Delta s\,b_{2},
\label{Eq10}
\end{equation}
where $\Delta s$ is the adjusted path length of the ray or the width of ionospheric layers. Note that at the new position $a_2, b_2$ become $a_1, b_1$, correspondingly. So, the direction cosines $a_2, b_2$ need to be calculated at each iteration step.

It is important to define correctly the direction cosines $a_1, b_1$ at the starting point. Recall that an incidence of a plane radio wave from the Sun into the ionosphere is under investigation. In the horizontal coordinate system, which is a topocentric system, position of the Sun is specified by the azimuth and the elevation angle $\theta$. Only the latter will be used for the present two-dimensional problem. Assuming spherically stratified ionosphere, the elevation angle $\theta_{I}$ at a height $h_I$ can be expressed using Equation~(4) from \citet{Birch2002} to relate $\theta$ and $\theta_{I}$:
\begin{equation}
\cos\theta_{I} = \frac{\cos\theta}{(1+h_I/R_E)},
\label{Eq11}
\end{equation}
where $R_E$ is the radius of the Earth, 6371 km, and $h_I$ is the ``effective height'' of ionosphere. The numerical quantity of $h_I$ will be given below. Therefore, value of $\cos\theta_{I}$ is the directional cosine $a_1$, while $b_1$ can be determined as $b_1 = \sqrt{(1-a_1^2)}$. The plane wave will be represented by a set of radio rays with initial directional cosines $a_1$ and $b_1$.

\section{Simulation Results}
\label{Sect:Simulation Results}

To simulate the day-time MSTIDs, we have considered the typical parameters of a gravity wave having a horizontal wavelength $\lambda_{x}$ of 300 km, temporal period $T$ of 40 min (i.e. the phase velocity $c$ is equal to 125 m/s) and propagating in south-east direction with azimuth $A$ of 135$^\circ$ from the north. The dip angle $I$ is assigned to 63.5$^\circ$. The projection of the neutral wind velocity $u_{||}$ along the geomagnetic field is fixed to 7 m/s within the range of altitudes. The background wind velocity $U$ is assumed to be zero. The above-mentioned characteristic parameters have been taken from \citet{Otsuka2013}, where the authors performed calculations of MSTIDs over Europe.

The ambient electron density profile given by Equation~(\ref{Eq2}) is specified by peak electron density $N_{\textrm{m}}$, height of peak electron density $z_{\textrm{m}}$, and scale height $H$, which are set to be $1.24\cdot10^{12}$ m$^{-3}$, 300 km, and 50 km, respectively. Further, taking typical value of the Brunt-V$\"{a}$is$\"{a}$l$\"{a}$ frequency $\omega_{b} = 2\pi/(12\cdot60)$ s$^{-1}$, from Equation~(\ref{Eq3}) we can compute the vertical component of the wave number $k_{z}$. Before proceeding to Equation~(\ref{Eq4}), it should be noted that according to the World Magnetic Model for Epoch 2015 the magnetic declination $\delta$ slightly varies from $-2^{\circ}$ to $+8^{\circ}$ over Western, Central, and partially Eastern European regions (https://www.ngdc.noaa.gov/ngdc.html). Here we set $\delta$ to be 0$^{\circ}$. All the above presented values should be put into Equation~(\ref{Eq5}).

\begin{figure}[H]
\begin{center}
\includegraphics[width=0.95\textwidth]{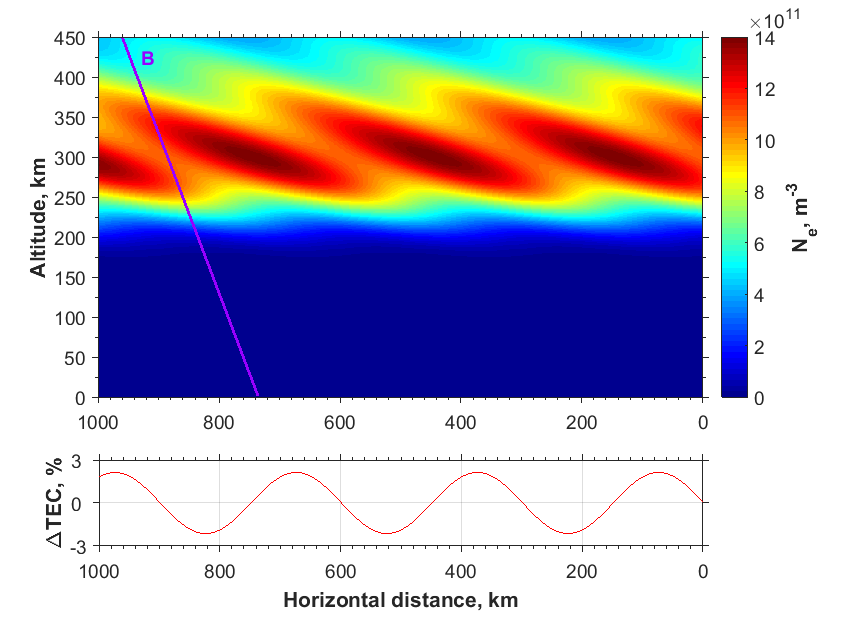}
\caption{Model calculation of MSTID structures produced by an atmospheric gravity wave with spatial and temporal periods of 300 km and 40 min, respectively. The gravity wave propagates equatorward with azimuth of $135^{\circ}$ (from the north). The AGW/TID movement is from right to left. The purple line in the top panel indicates the dip angle of the magnetic field that is equal to $63.5^{\circ}$. The bottom panel demonstrates the relative amplitude of TEC disturbances in terms of percentage. The vertical TEC is calculated by integration of the electron density within 250-400 km range of altitudes.}
\label{Figure1}
\end{center}
\end{figure}

The result of our model calculation is shown in Figure~\ref{Figure1}. It is seen that the MSTID pattern represents a periodic sequence of enhancements and depressions of electron concentration. Besides, the TID front in the vertical plane inclines towards the direction of its propagation. In the figure, the TID movement is from right to left. This configuration is significant with respect to the TID focusing effect and will be discussed below. To calculate the vertical total electron content (TEC), the electron density profiles were integrated within a layer of 150 km width, from 250 km to 400 km. The bottom panel in Figure~\ref{Figure1} displays the relative amplitude of TEC perturbations presented as $\Delta\textrm{TEC}=(T_1-T_0)/T_0\cdot100\%$, where $T_1$ and $T_0$ are the vertical TEC with and without the gravity wave, respectively. The maximum of $\Delta\textrm{TEC}$ takes a value of $2.13\%$ for $\gamma = 1$ (see Equation~(\ref{Eq5})) that is consistent with the observed TEC perturbations associated with MSTIDs.

\begin{figure}[H]
\begin{center}
\includegraphics[width=1\textwidth]{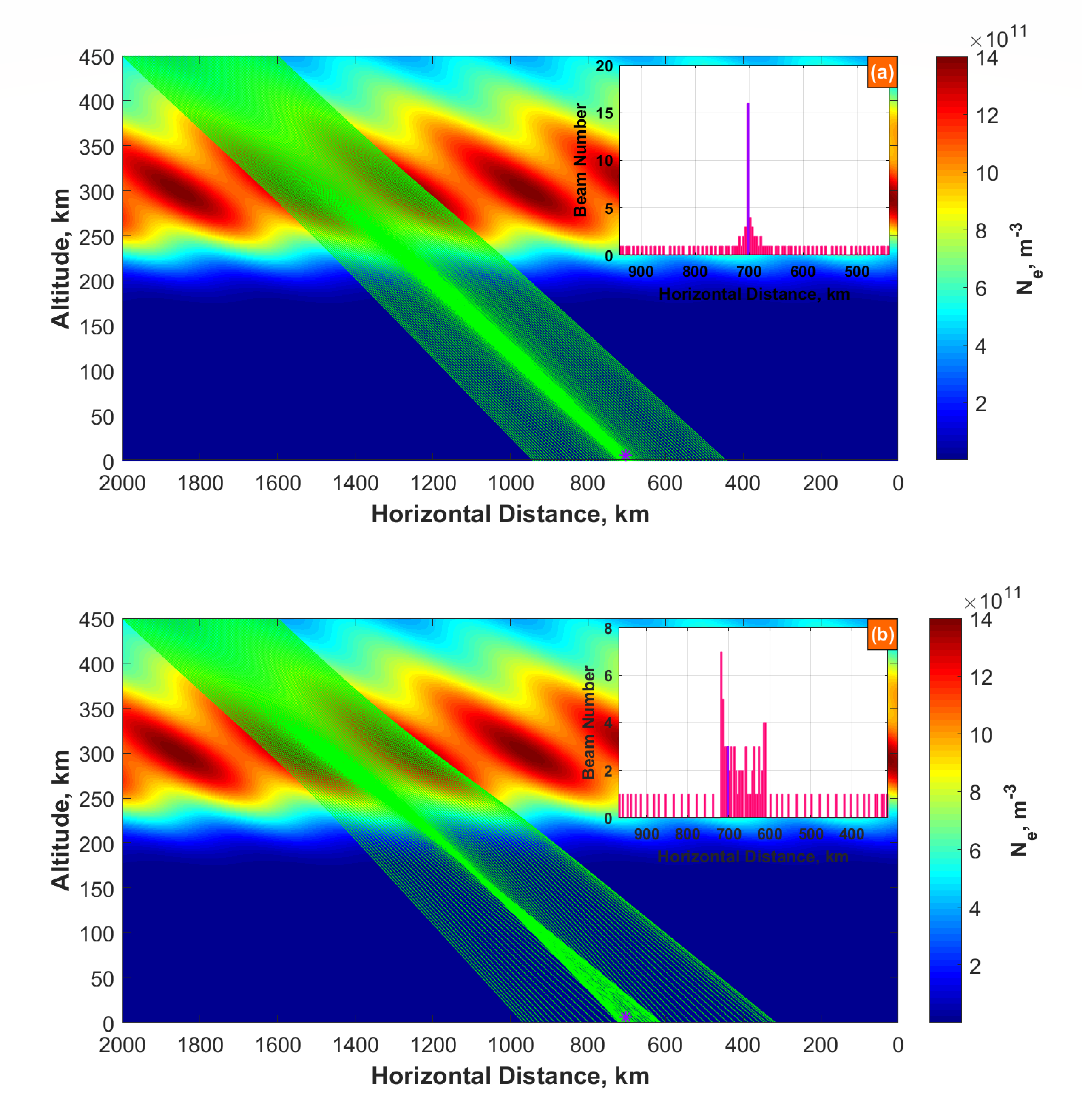}
\caption{Examples of the simulations of the propagation of radio rays (green lines) at 105 MHz (a) and 75 MHz (b) through the perturbed ionosphere with the same parameters as for Figure~\ref{Figure1}. The rays are uniformly distributed in the range of distances from 1600 to 2000 km with 4 km spacing. The starting point ($h_I$) in altitude is 450 km. The elevation angle ($\theta$) is equal to $8^{\circ}$. The histograms in both panels demonstrate the number of beams falling into the 4-km distance at the ground level. The histogram bin width is 4 km.}
\label{Figure2}
\end{center}
\end{figure}

In Figure~\ref{Figure2}, two representative examples of refraction of radio waves (radio rays) at frequencies 105 MHz (a) and 75 MHz (b) in the modelled ionosphere are shown. The rays come out from points distributed between 1600 and 2000 km with a 4-km step along horizontal distance and placed at 450 km in altitude. The angle of incidence of all rays at the starting points $\theta_{I}$ is recalculated using Equation~(\ref{Eq11}) for the case of solar elevation angle $\theta = 8^{\circ}$ and the ``effective height'' $h_I = 450$ km. Besides, we set $\delta x = \delta y$ = 1 km (see Equation~(\ref{Eq7}),~(\ref{Eq9})), and the path length $\Delta s$ to 1 km (see Equation~(\ref{Eq10})).

The ray trajectories are determined by the ray-tracing algorithm as described in Section~\ref{Sect:Ray tracing method}. Note that panels (a) and (b) of Figure~\ref{Figure2} are plotted for demonstrative purpose to make the simulation process easily understandable. Each panel presents a picture of radio rays at the same instant with the only difference in radio wave frequency. At a receiving point located at the ground level the number of incoming radio rays is counted. In the figure, the selected range of distances (cell) -- 700-704 km -- is marked by purple asterisk, while the purple histogram bars indicate the number of rays getting into this distance range. An increase of the number of radio beams in the cell up to 16 for 105 MHz and up to 3 for 75 MHz is registered.

\begin{figure}[H]
\begin{center}
\includegraphics[width=0.95\textwidth]{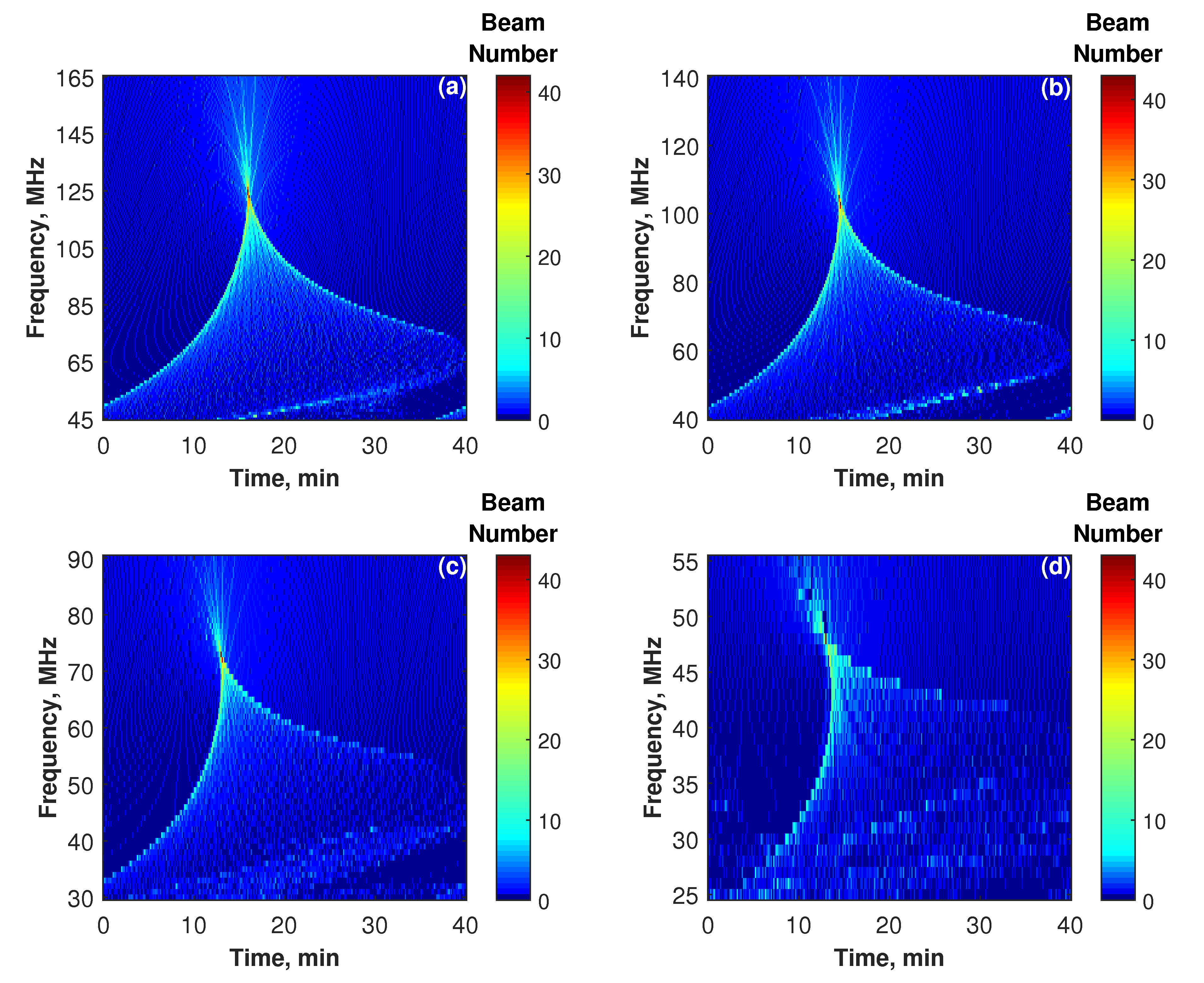}
\caption{The beam-intensity in the time-frequency plane (i.e. dynamic spectrum) obtained by counting numbers of radio rays received in the fixed 1-km distance at the Earth's surface (at the presumed observational site). The simulation is performed with 1 MHz resolution in frequency and 2/15 min resolution in time.  The dynamic spectra were produced under different solar elevation angles $\theta$: (a) $2^{\circ}$, (b) $8^{\circ}$, (c) $14^{\circ}$, (d) $20^{\circ}$. The color scale indicates the number of beams recorded at the presumed site of observation.}
\label{Figure3}
\end{center}
\end{figure}

Figure~\ref{Figure3} shows the main result of our calculations. Here the beam density has been increased by reducing the beam spacing to 1 km. Therefore, the number of incoming beams every 1-km distance at the ground surface was counted, while the propagation of TIDs with the spatial period of 300 km is simulated by moving the structures every 1/300 of the temporal period $T$, i.e. 40/300 min = 2/15 min. At the same time by changing the frequency of radio rays with a 1-MHz incremental step, we recorded the beam intensity in the time-frequency domain. In such a way we simulated the solar dynamic spectra for several elevation angles $\theta$ ($2^{\circ},8^{\circ},14^{\circ},20^{\circ}$). In each dynamic spectrum there is a point with the highest beam-intensity corresponding to the focusing frequency. We have found the dependence of the focusing frequency on the elevation angle $\theta$ for its values from $2^{\circ}$ to $30^{\circ}$ (see Figure~\ref{Figure5}), as will be discussed in the next section.

\section{Discussion}
\label{Sect:Discussions}

The main results of the study are partially presented in Figure~\ref{Figure3}. It shows four dynamic spectra corresponding to different values of solar elevation angle ($\theta = 2^{\circ},8^{\circ},14^{\circ},20^{\circ}$). Each dynamic spectrum includes distinctive spectral perturbation in intensity that can be recognized as a SC. This is consistent with the established result that the focusing effect of TIDs during solar radio observations can produce similar spectral structures.

\begin{figure}[H]
\begin{center}
\includegraphics[width=1\textwidth]{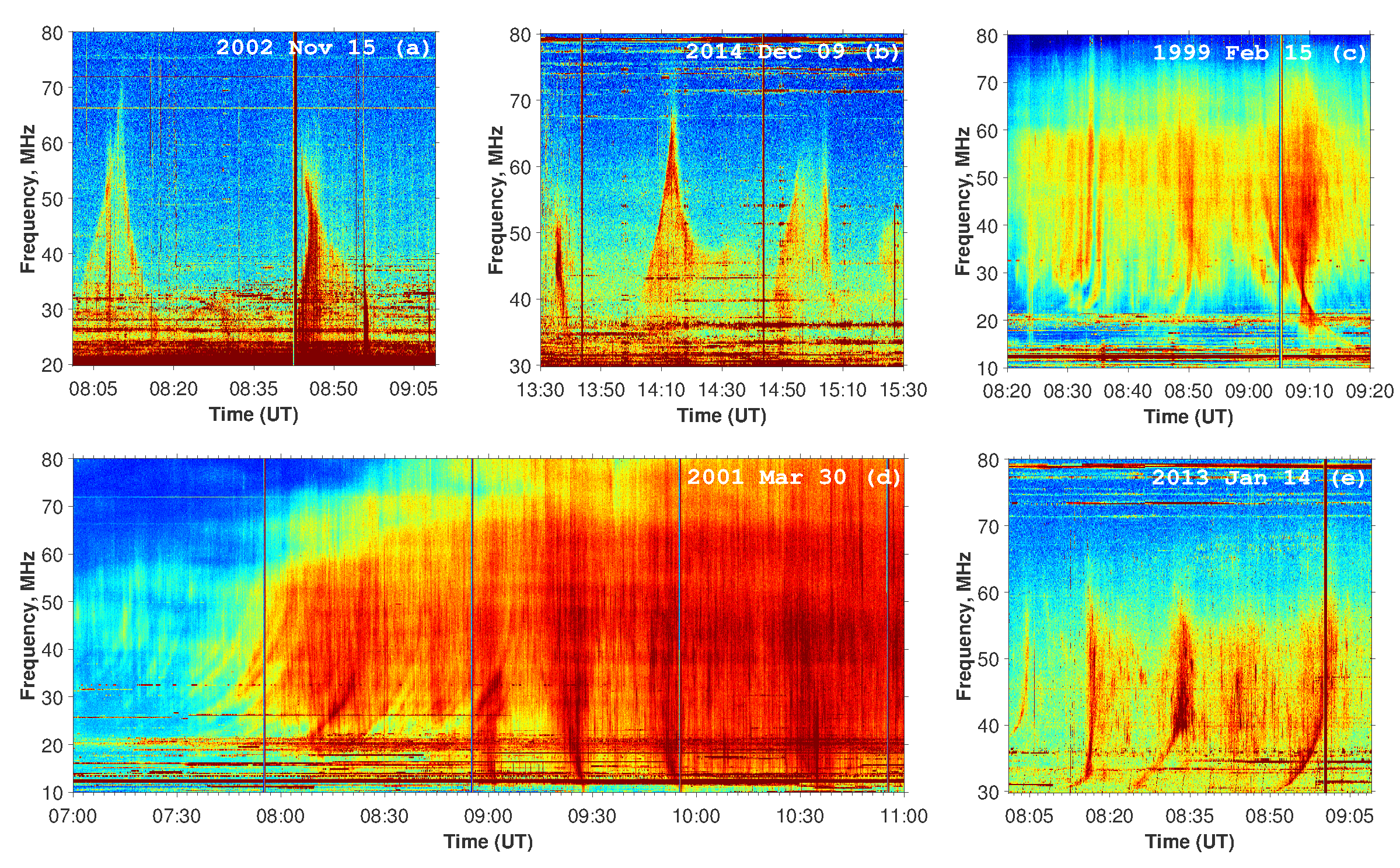}
\caption{Examples of the NDA solar dynamic spectra including different SC types: (a and b) inverted V like, (a and c) X like, (d) V like, (e) fiber like, and (c and d) fringe
like. Some spectrograms contain multiple SCs belonging to separate types (see description in the text). The dynamic spectra are credited from the paper by \citet{Koval2017}.}
\label{Figure4}
\end{center}
\end{figure}

To show direct comparison of the simulation results with the observed SCs, we have introduced Figure~\ref{Figure4}. It represents the set of solar dynamic spectra with SCs obtained by the NDA. Note that this figure has been assembled by using some spectrograms from Figure~1 of our earlier paper \citep{Koval2017}. That work is entirely devoted to the SCs observations based on the NDA solar emission measurements. There we have divided all SCs into several types: inverted V-like, V-like, X-like, fiber-like and fringe-like. The first one - inverted V-like type - prevails over other SC types. To make the present paper more solid as well as to confirm reliability of the preformed simulations, the actual SCs have been shown. More observational examples can be found in our previous paper on the issue.

From Figure~\ref{Figure3}, it is seen that a typical SC structure consists of front and back envelopes and a body between them. The envelopes have higher brightness than the interior and approach each other at a certain convergent point that is characterized by the peak brightness of the whole structure. The visible broadening from the convergent point occurs with decreasing frequency that makes the structure wedge-shaped. This description agrees well with the SC definition of inverted V-like type which has been introduced together with other types of SCs in our recent paper \citep{Koval2017}. This type of SC can be seen in Figure~\ref{Figure4}(a) near 08:10 UT, in Figure~\ref{Figure4}(b) near 14:10 UT and 15:00 UT.

The frequency of the convergent point is the focusing frequency. It implies that with current parameters of the ionosphere and solar radiation a ground observer is in the focus of a plasma lens formed by TIDs. In Figure~\ref{Figure3}(a-d) it happens at frequencies of 125~MHz, 105~MHz, 73~MHz, 48~MHz, respectively. With decreasing frequency the focusing point moves higher in altitude like in Figure~\ref{Figure2}(b), where it is at a height of approximately 160 km. With increasing frequency the focusing point is located below the ground level and therefore not observable. For dynamic spectra recorded above the focusing frequencies, the caustic structure manifests as a fan of straight, separate stripes diverging to high frequencies. A dynamic spectrum including the focusing frequency demonstrates the structure in the shape of the letter X, yet with the upper half smaller than the lower half (see Figure~\ref{Figure3}(a-c)). We believe that the observed X-like type of SCs (see Figure~\ref{Figure4}(a, c) near 08:50 UT and 09:10 UT, respectively) is also well reproduced by our simulations. Possibly, for some events the upper half of the X-like SCs may get too week in intensity to be detected. This yields the inverted V-like SC.

It is interesting to note that the X-like type of the SCs could turn into another type of SCs, namely V-like type. In solar spectral observations, the latter takes wedge-shaped form with broadening to high frequencies. We suggest that in these cases only the upper part of the X-shaped pattern is detected. Note that the V-like type SCs are rather rare events which appear at low frequencies ($\sim$10-30 MHz) in the NDA dynamic spectra (see Figure~\ref{Figure4}(d) around 09:00 UT, 09:30 UT and 10:00 UT). It appears that the bottom part of the original X-like SC lies in the lower boundary of the NDA working band. Presumably, the radio emission received by NDA below 30 MHz may be attenuated by gain-frequency characteristic of the antennas at these lower frequencies. Another possible explanation of cutting a SC of X-like type may be related to some peculiar solar and ionospheric parameters during the formation process of SCs.

According to panels (a-d) of Figure~\ref{Figure3}, a SC pattern presents an anticlockwise rotation as a whole with increasing elevation angle. This behavior is associated with the relation between the front slope of the TID and the angle of incidence of the radio wave. Usually the TID front in the vertical plane inclines towards its moving direction. The inclination $\xi$ is caused by the difference in phase velocities of a gravity wave at lower and upper boundaries of the slab with TIDs \citep{Fedorenko2013}. It can be calculated as the ratio of the vertical wave number $k_{z}$ to the horizontal wave number $k_{x}$ in the form of $\xi = \arctan(k_{z}/k_{x})$ (see Equation~(\ref{Eq4})). In the MSTID model used in this work, $\xi$ is found to be $72.35^{\circ}$. Further, for $\theta = 2^{\circ}$ the angle of incidence of radio rays in the vertical plane can be expressed as $(90^{\circ}-\theta_I)$ that is equal to $68.98^{\circ}$. It means that the SC in Figure~\ref{Figure3}(a) is generated under incidence of radio wave that is quasi-normal to plasma density irregularity. That is why the SC in panel (a) presents a nearly symmetric pattern.

From Figure~\ref{Figure3}(b-d) it is seen that the front envelope becomes steeper with solar elevation angle, while the back envelope grows flatter and diffusive. For $\theta = 20^{\circ}$ the back envelope of the SC is hardly recognized, while the front envelope developed a thread-like structure with a rapid drift in frequency. In our classification of SC types, the SCs of fiber-like type have similar spectral morphology. They emerge in forms of bright filaments and/or lanes with no frequency broadening/narrowing. Thus, we suggest that the fiber-like type SCs are actually the front envelope of a whole SC's pattern that is only partially observed. This suggestion seems to be feasible based on comparison of the simulations with real fiber-like SCs in Figure~\ref{Figure4}(e) near 08:15 UT, 08:30 UT and 09:00 UT.

\begin{figure}
\begin{center}
\includegraphics[scale=0.5]{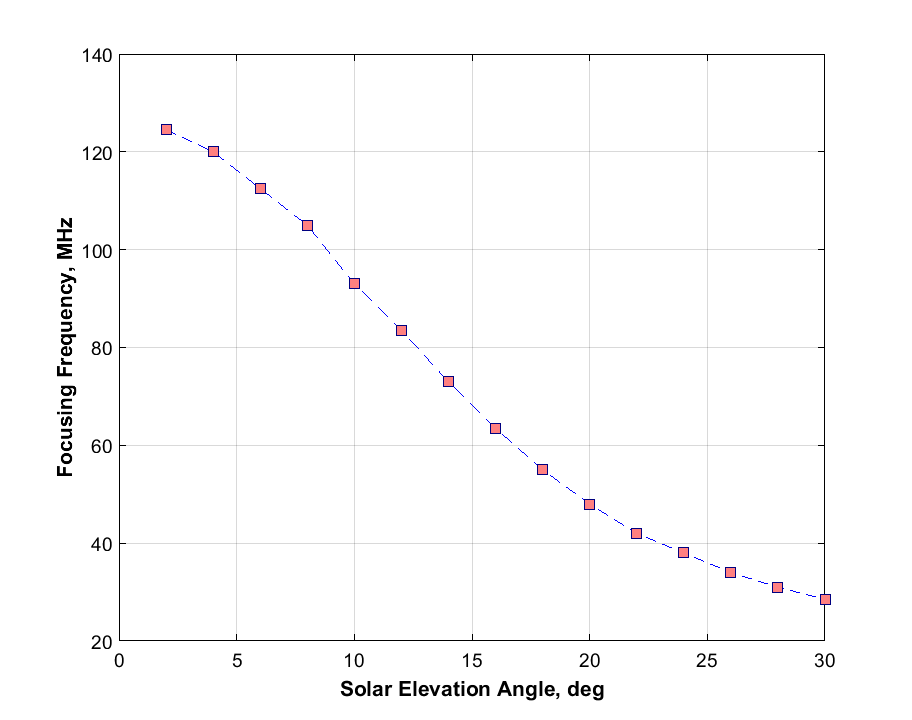}
\caption{Dependency of the focusing frequency on the elevation angle of the Sun. The values of the focusing frequency (orange squares) are determined every $2^{\circ}$.}
\label{Figure5}
\end{center}
\end{figure}

The dependency of the focusing frequency on solar elevation angle is presented in Figure~\ref{Figure5}. The point with the focusing frequency has the maximum intensity in a dynamic spectrum. The largest portion of a SC's body lies lower than the focusing frequency (see Figure~\ref{Figure3}). This characteristic parameter indicates where the main part of a SC would be located in the solar dynamic spectrum. We point out that the relation on Figure~\ref{Figure5} has been obtained only for present configuration of the ionosphere (regular and disturbed parts) and in the range of the elevation angles $\theta$ from $2^{\circ}$ to $30^{\circ}$.

Figure~\ref{Figure5} shows that the focusing frequency rapidly falls with the growth of the elevation angle. The low values of $\theta$ correspond to typical positions of the Sun in winter and partially in spring and autumn months at middle latitudes of Europe. The maximum focusing frequency is 125 MHz for $\theta = 2^{\circ}$. This may be considered to support the fact that the SCs can spread up to 200 MHz that was established from solar observations. At the elevation angle equal to $25^{\circ}$ the focusing frequency is below 40 MHz. Based on the simulation result in Figure~\ref{Figure3}(d) for $\theta = 20^{\circ}$, a SC at larger $\theta$ would be partially or completely damaged, or not be generated at all. Thus, we infer that the SCs may be observed only in certain periods, mainly in late fall, winter, and early spring. This selectivity in generations of SCs explains the seasonal dependence in the SC appearance which we found in the NDA data \citep{Koval2017}. In that paper we have attributed such behavior to similar seasonal variations in the TID occurrence rate \citep{Hernandez2006,Tsugawa2007,Otsuka2013}. It is difficult to conclude unambiguously whether both factors play a role or the two factors are actually correlated with each other.

Concerning the inclination angle $\xi$ of the TID front, the following should be noted. As we mentioned, $\xi$ is a constant in our simulation. In reality it varies with the distance $R$ from the TID source. \citet{Fedorenko2013} have presented the relation between $\xi$ and $R$. The authors showed that the inclination grows fast, almost linearly, in a range of distances from a TID emitter to about 2000 km, where $\xi$ is near $75^{\circ}$. With a further increase in the distance, the inclination rises slowly and reaches $85^{\circ}$ at a length of 8000 km. Apparently, the smaller values of the inclination $\xi$, i.e. when a TID is at relatively short distances away from its source, will result in the lower focusing frequency of a SC in the dynamic spectrum. Consequently, one should realize that the positional relationship of the TID front and a plane wave is crucial to define the morphological and spectral characteristics of SCs.

\section{Conclusions}
\label{Sect:Conclusions}

In our previous paper on the focusing effect of solar radio emission by the Earth's ionosphere \citep{Koval2017}, we have performed long-term statistical analysis and classification of the SCs using the NDA observational data. In this study, we move from the observational analysis to a modelling effort. We applied the ray tracing technique to compute radio beam trajectories in the terrestrial ionosphere, including the medium scale TIDs. The dynamic spectra with structures consistent with the observed SCs have been obtained. With the simulation, we are able to identify four types of SCs among the five ones declared by our earlier study, including the inverted V-like, V-like, X-like, and fiber-like types. This proves, firstly, the reliability of the introduced classification of the SCs; secondly, the correct numerical treatment of the issue; thirdly, further studies are required for explanation of the last type of SCs, i.e., the fringe-like type.

The fringe-like type of SCs has a look of alternating stripes which are enhanced and depressed in intensity (see Figure~\ref{Figure4}(c) within 08:20 UT -- 09:00 UT, and Figure~\ref{Figure4}(d) within 07:00 UT -- 08:30 UT). It should be noted that the similar morphology to the fringe-like one appears as the fine structure of some decameter type II bursts in the form of drifting narrowband fibres on the dynamic spectrum \citep{Afanasiev2009}. It is generally accepted that the type II bursts are associated with shock waves, which can be produced by coronal mass ejections (CMEs). The author suggests that the fibres are manifestations of caustics, which are generated due to refraction of radio waves on the inhomogeneous density structure of the CME front. This concept can be adapted for our case. Probably, if the internal structure of a TID cell has electron density variations, it may give rise to multiple caustics that will show up in the form of fringe-like intensity patterns on the dynamic spectrum.

Based on the simulations, we found the dependence of the focusing frequency on the elevation angle of the Sun. The focusing frequency belongs to a characteristic point in the SC structure which has a peak intensity. The focusing frequency can be recorded when a ground observer is near the focal point of a plasma lens created by a TID. Hence, using the established relation it is possible to predict a frequency range within which a SC may appear. We revealed that the SCs may be recorded in spectrograms for certain elevation angles of the Sun. At relatively low solar elevation angles (<$25^{\circ}$), the SCs can be generated. This range of elevation angles corresponds to late fall, winter, and early spring. This provides a nice explanation of the seasonal dependence in SC occurrence which has been noticed in our previous paper. According to the latter, 95 $\%$ of days with SCs belong to autumn-winter months. Moreover, we found a close correlation between the slope of the TID front and the angle of incidence of solar radiation through the ionosphere. This correlation affects spectral and morphological properties of SCs. In particular, the pattern of a SC is mostly symmetric when the angle of incidence and the TID inclination angle are close to each other. If there is a considerable difference between them, the pattern loses its symmetry with the modification of its envelopes and rotation in solar radio spectrogram.

We should note that our study is motivated by the observational analysis of the data received by the NDA antenna. This instrument is located in Nan\c{c}ay (France) with coordinates: $47^{\circ}23'$ N, $2^{\circ}12'$ E. Thus, the parameters of the ionospheric model and solar emission have been set to those relevant to middle latitudes of northern hemisphere. It is expected that as long as the ionosphere reveals similar properties of TIDs as those modelled here, similar behavior of SCs can be observed at other locations of the Earth. Along with that, we point out that the model of the regular ionosphere with fixed parameters, namely critical frequency of $F2$ layer $f_0F2$, has been accepted. Seasonal, daily, and day-to-day variations, as well as the trend of $f_0F2$ depending on the solar cycle are not considered in this work, that can be taken into account in further research of SCs.

\acknowledgments
This research was supported by NNSFC grants 41331068, 11790303 (11790300), 41774180, and NSBRSF 2012CB825601. The authors are thankful to Dr. Yuichi Otsuka for valuable consultations about the MSTID model. The data files and software for plotting SCs as well as the focusing frequency dependence are accessible at https://figshare.com/  (DOI: 10.6084/m9.figshare.6833312).




\listofchanges

\end{document}